\documentclass[10pt]{iopart}
\usepackage{graphicx}
\usepackage{subfigure}
\usepackage{verbatim}
\usepackage{dcolumn}
\usepackage{bm}
\usepackage{epsf}
\usepackage{color}
\usepackage[toc]{appendix}
\usepackage{stmaryrd}
\usepackage{amsthm}
\usepackage{hhline}
\usepackage{amssymb}
\usepackage{iopams}
\usepackage{cite}
\usepackage{tikz}
\usetikzlibrary{arrows,shapes,calc,matrix}

\expandafter\let\csname equation*\endcsname\relax
\expandafter\let\csname endequation*\endcsname\relax
\usepackage{amsmath}
\usepackage{xstring}

\makeatletter
\newcommand\footnoteref[1]{\protected@xdef\@thefnmark{\ref{#1}}\@footnotemark}
\makeatother

\newcommand\myeqref[1]{
	Eq. (\textup{\ref{#1}})
}
\newcommand\invisiblesection[1]{%
	\refstepcounter{section}%
	\addcontentsline{toc}{section}{\protect\numberline{\thesection}#1}%
	\sectionmark{#1}}
\newcommand{\bra}[1]{\left\langle #1\right|}
\newcommand{\ket}[1]{\left|#1\right\rangle}

\newcommand{\trace}[1]{\mathrm{tr}\left\{#1\right\}}

\newcommand{\la}{\left\langle}
\newcommand{\ra}{\right\rangle}

\newcommand{\mi}[1]{\min{\left\{#1\right\}}}

\newcommand{\bla}{bla\\bla\\bla\\bla\\bla}
\usepackage{fmtcount}

\newcommand{\mc}[1]{\mathcal{#1}}

\newcommand{\mrm}[1]{\mathrm{#1}}

\renewcommand{\appendix}{
}

\newcommand{\draftmode}{1}    
\newcommand{\notetoself}[1]{\ifnum \draftmode=1 {\color[rgb]{0,0,0.8} [#1]} \fi}  
\newcommand{\cuttext}[1]{\ifnum \draftmode=1 {\color[rgb]{0,0.5,0} [#1]} \fi}  
\newcommand{\warntext}[1]{\ifnum \draftmode=1 {\color[rgb]{0.9,0.6,0} #1} \else {#1} \color{black} \fi}
\newcommand{\aref}[1]{{Appendix~\hyperref[#1]{A}}}
\newcommand{\bref}[1]{{Appendix~\hyperref[#1]{B}}}
\usepackage[colorlinks=true,citecolor=blue,linkcolor=blue,urlcolor=blue]{hyperref}
\usepackage{doi}
\usepackage{cleveref}

\begin{document}

\title{Quantum scrambling and the growth of mutual information}

\author{Akram Touil\textsuperscript{1} and Sebastian Deffner\textsuperscript{2}}
\address{Department of Physics, University of Maryland, Baltimore County, Baltimore, MD 21250, USA}
\ead{$^1$akramt1@umbc.edu}
\ead{$^2$deffner@umbc.edu}

\begin{abstract}
Quantum information scrambling refers to the loss of local recoverability of quantum information, which has found widespread attention from high energy physics to quantum computing. In the present analysis we propose a possible starting point for the development of a comprehensive framework for the thermodynamics of scrambling. To this end, we prove that the growth of entanglement as quantified by the mutual information is lower bounded by the time-dependent change of Out-Of-Time-Ordered Correlator. We further show that the rate of increase of the mutual information can be upper bounded by the sum of local entropy productions, and the exchange entropy arising from the flow of information between separate partitions of a quantum system. Our results are illustrated for the ion trap system, that was recently used to verify information scrambling in an experiment, and for the Sachdev-Ye-Kitaev model.
\end{abstract}

\section{Introduction}

One of the most intriguing problems in theoretical physics is the information paradox \cite{hawking1976black,preskill1992black,Mathur2009}, which suggests that physical information crossing the event horizon could permanently disappear in a black hole. In essence, the paradox originates in the unresolved incompatibility of current formulations of quantum mechanics and general relativity. To date, many possible solutions have been proposed, some as esoteric as the many-worlds interpretation of reality \cite{preskill1992black,poplawski2010cosmology}, whereas others are rooted in quantum information theory \cite{Kallosh2006,Giddings2012}.

A particularly fruitful concept has been dubbed \emph{quantum information scrambling} \cite{swingle2018unscrambling}. Within this paradigm, information that passes the event horizon is quickly and chaotically ``scrambled'' across the entirety of the horizon. Thus, the information only appears lost, as no \emph{local} measurement allows to fully reconstruct the original quantum state \cite{hashimoto2017out,Sekino_2008,Hayden_2007}. In recent years, the study of quantum information scrambling has led to new physical concepts, such as the black hole complementarity and the holographical principle \cite{susskind1993stretched,susskind1995world}. In addition, information scrambling has found attention in high energy physics \cite{Ling2017,grozdanov2018black}, quantum information \cite{couch2019speed,Yoshida2019}, condensed matter physics \cite{blake2018many,Iyoda2018}, and quantum thermodynamics \cite{campisi2017thermodynamics,Yunger2017,Chenu2019workstatistics}. 

Remarkably, it has also been recognized that exploiting the AdS/CFT duality \cite{maldacena1999large} and the  ``ER=EPR-conjecture" \cite{maldacena2013cool},  the information scrambling dynamics of black holes can be studied with analog quantum systems. Loosely speaking, the dynamics of two black holes connected through an Einstein-Rosen bridge can be mathematically mapped onto the dynamics of entangled quantum systems. In fact, this idea led to the first verification of quantum information scrambling \cite{landsman2019verified}  in an ion trap experiment.

This ubiquity of information scrambling poses the question which underlying physical principles determine, if, when, and how information is distributed. Since information is physical \cite{landauer1961irreversibility}, and its processing requires thermodynamic resources \cite{bennett1973logical,bennett1982thermodynamics,zurek1984reversibility,deffner2013information,Parrondo2015,Boyd2016}, it is only natural to realize that the second law of (quantum) thermodynamics \cite{Deffner2019book} must hold the answer. The task is then to uniquely quantify the thermodynamics resources (such as heat or work) that are consumed while information is scrambled.

In the literature, a plethora of quantifiers have been proposed that can track and characterize quantum information scrambling, such as, for instance, Out-of-Time-Ordered Correlators (OTOCs) \cite{maldacena2016bound,swingle2016measuring,nakamura2019universal,bergamasco2019otoc}, the Loschmidt Echo \cite{yan2019information,Sanchez2020}, and versions of the mutual information \cite{iyoda2018scrambling,seshadri2018tripartite,Alba2019}. Out of  this variety of measures, the OTOC has probably gained the most attention. This is due to the fact that the properties of the OTOC characterize the dynamical emergence of ``non conventional'' quantum chaos \cite{Roberts2016,iyoda2018scrambling}. To the very best of our knowledge, however, a concise, transparent, and practically relevant relationship between the OTOC and a thermodynamic observable appears to be lacking \footnote{For obvious reasons, the projective measurements considered in Ref.~\cite{campisi2017thermodynamics} are neither feasible nor practical in complex many body systems}. 

Therefore, in the present analysis we prove upper and lower bounds on the quantum mutual information. As main results, we find (i) that the time-dependent mutual information is lower bounded by the change of the OTOC, and (ii) that the rate of change of the mutual information is upper bounded by the sum of the stochastic entropy productions \cite{esposito2010three} in the separate partitions of a quantum system.  Our findings are illustrated for the experimental system described in Ref.~\cite{landsman2019verified} and an example of a quantum chaotic system, the SYK model.
	
\section{Time-dependent mutual information and the change of the OTOC}
\label{sec2}

Imagine a quantum system $S$ that can be separated into two partitions, $A$ and $B$. The total system $S$ evolves under unitary dynamics, and the quantum state is initially prepared as a product, $\rho_S(0)=\rho_A(0)\otimes \rho_B(0)$.  Typically, quantum information scrambling then occurs in situations, in which $\rho_S(0)$ is chosen to be pure, and the unitary dynamics of $S$ yields the growth of entanglement between $A$ and $B$. This means, in particular, state tomography on  only $A$ is not longer sufficient to reconstruct $\rho_A(0) $ for any time $t>0$.

The loss of local recoverability is conveniently characterized by the Out-of-Time-Ordered-Correlator (OTOC) \cite{swingle2018unscrambling,hashimoto2017out,Sekino_2008,Hayden_2007}, which can be written as,
\begin{equation}
\mathcal{O}(t)=\la O_{A}^{\dagger} O_{B}^{\dagger}(t)\, O_{A} O_{B}(t)\ra\,.
\label{1}
\end{equation}
Here, $O_{A}$ and $O_{B}$ are  \emph{local} operators acting only on $A$ and $B$, respectively. More specifically, we have $O_{A}\equiv o_{A} \otimes I_{B}$, $O_{B}\equiv I_{A} \otimes o_{B}$, and $O_{B}(t)= U^{\dagger}(t)\,O_{B}(0)\,U(t)$, where $U(t)$ denotes the unitary time evolution operator of $S$. It has been argued, that $\mc{O}(t)$ characterizes the spread of the operator $O_{B}(t)$ as it evolves in time, which tracks how information is scrambled from $A$ to $B$ \cite{swingle2018unscrambling,hashimoto2017out,Sekino_2008,Hayden_2007}. The average in Eq.~\eqref{1} is often taken over a thermal state in $S$ \cite{swingle2018unscrambling,yan2019information}, which is, however, not necessarily an instrumental choice \cite{yan2019information}.

Note, however, that (to the best of our knowledge) there is no rigorous proof of the existence of operators correctly tracking information scrambling in any physical scenario. For instance, taking operators that commute with the Hamiltonian of $S$ results in a time-invariant OTOC. Thus, one often takes an average over operators, rather then working with $\mathcal{O}(t)$ directly \cite{cotler2017chaos,roberts2017chaos}.

\subsection{Lower bound on quantum mutual information}

For the following analysis, we will be motivated by the conceptual framework that maps notions from high energy physics onto a quantum information theoretic language \cite{maldacena2013cool}.
\begin{figure}
\centering
\includegraphics[width=0.65\textwidth]{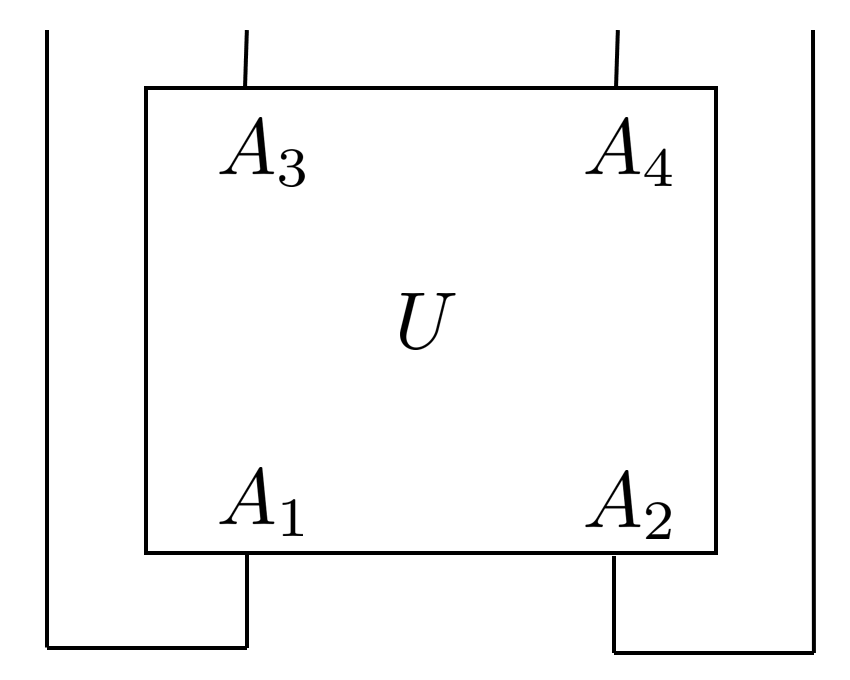}
\caption{Sketch of a black hole scrambling quantum information. $A_1$ characterizes the information falling into the black hole, $A_2$ determines the initial state of the black hole, $A_4$ the Hawking radiation, and $A_3$ the remaining black hole.}
\label{f2}
\end{figure}
To this end, imagine a situation in which some quantum information falls across the event horizon of a black hole, and we can describe the dynamics of that black hole by a scrambling (entangling) unitary map, $U(t)$. The set-up is depicted in Fig.~\ref{f2}, which is similar in spirit to Ref.~\cite{Hosur2016}. Hosur \etal consider that $(A_1, A_3)$ and $(A_2, A_4)$ describe EPR pairs, connected through the scrambling unitary $U$. More specifically, $A_1$ determines the initial information thrown into the black hole, $A_2$ encodes the initial state of the black hole, $A_4$ the Hawking radiation and $A_3$ the remainder of the black hole.

For such scenarios, it has been shown \cite{Fan2017,Li2017,hunter2018chaos} that
\begin{equation}
\label{eq:inequ}
\la O_{A_1}O_{A_4}(t)\,O_{A_1}O_{A_4}(t)\ra_{\mrm{avg}}=2^{-\mathcal{I}^{(2)}_{A_1, A_2A_4}}\,,
\end{equation}
where $\mathcal{I}^{(2)}_{i, j}$ is the R\'{e}nyi-2 mutual information between the partitions $i$ and $j$. Notice that $A_2$ appears in the subscript of $\mathcal{I}^{(2)}$, which is related to why knowing $A_2$ and $A_4$ specifies $A_1$. In the case of a unitary map modeling the scrambling of information in a black hole the Hawking radiation ($A_4$) and knowledge of the initial state of the black hole ($A_2$) is enough to reconstruct the quantum state described by $A_1$. This can be seen by realizing that black holes can be regarded as quantum error correcting codes, i.e., if we wait long enough, whatever information that got scrambled within the black hole can be accessed through only Hawking radiation and the initial state of the black hole (that is we do need to know the state of the black hole before any information was thrown into it), without the need to access any other degrees of freedom \cite{preskill1992black}.

For initially pure states\footnote{For the sake of simplicity we only consider initially pure states. If the composite quantum state was initially mixed, we would obtain the same results up to additive constants.}, we can write
\begin{equation}
\mathcal{I}^{(2)}_{i,j} = {\mc{S}_i}^{(2)}+{\mc{S}_j}^{(2)} = -\ln(\trace{\rho^{2}_i})-\ln(\trace{\rho^{2}_j})\,.
\end{equation}
Furthermore, the average in Eq.~\eqref{eq:inequ} is taken over the Haar measure on the unitary group U(d) with
\begin{equation}
\int_{\mathrm{Haar}} d U=1,
\end{equation}
and where we have for an arbitrary function $f$ and $\forall \ V \in U(d)$,
\begin{equation}
\int_{\text {Haar }} d U f(U)=\int_{\text {Haar }} d U f(V U)=\int_{\text {Haar }} d U f(U V)\,.
\end{equation}
It is interesting to note that for quantum systems comprised of qubits, such as the experimental system analyzed in Ref.~\cite{landsman2019verified}, the Haar average is equivalent to an average over the Pauli group for each operator \cite{roberts2017chaos}.

Equation~\eqref{eq:inequ} can now be used to relate the OTOC with a thermodynamically relevant quantity, the quantum mutual information. Adopted to our current purposes $A_1$ and $A_3$ are operators that live on subsystem $A$, and $A_2$ and $A_4$ live on $B$. Therefore, we  consider
\begin{equation}
1-\la O_{A}O_{B}(t)\,O_{A}O_{B}(t)\ra_{\mrm{avg}}= 1-2^{-\mathcal{I}^{(2)}_{A, B}}\leq \mathcal{I}^{(2)}_{A, B}\,,
\end{equation}
where we identified $A_1\equiv O_A$ and $A_2 A_4\equiv O_B$. To justify this identification, we note that knowledge about the degrees of freedom $A_2$ and $A_4$ is enough to infer the information encoded in $A_1$. The same argument holds for any closed quantum system  and any unitary evolution $U(t)$. In this case, the analog of ``Hawking radiation'', is a subset of the degrees of freedom that is enough to reconstruct the initial information. 

Note that the R\'{e}nyi-2 mutual information  is upper bounded by the quantum mutual information $\mathcal{I}$ \cite{lieb1973eh,Verdu2015}. This is a direct consequence of the strong subadditivity of the von Neumann entropy \cite{lieb1973eh,Verdu2015}. We have
\begin{equation}
\label{eq:mut_info}
\mathcal{I}(t)=\mc{S}_{A}(t)+\mc{S}_{B}(t)-\mc{S}_{S}(t),
\end{equation}
where $S_{i}=-\trace{\rho_{i}\ln(\rho_{i})}$ is the von Neumann entropy of system $i$ with density matrix $\rho_{i}$. Note that $\mc{S}_{S}(t)=\mc{S}_{S}(0)$ for unitary dynamics. Thus, we immediately obtain
\begin{equation}
1-\la O_{A}O_{B}(t)\,O_{A}O_{B}(t)\ra_{\mrm{avg}} \leq \mathcal{I}(t)\,.
\end{equation}
Now introducing the notation $\bar{\mathcal{O}}(t)\equiv\la O_{A}O_{B}(t)\,O_{A}O_{B}(t)\ra_{\mrm{avg}} $ and noticing that by definition $\bar{\mathcal{O}}(0)= 1$, we can write
\begin{equation}
\label{eq:main1}
\mathcal{I}(t) \geq \bar{\mathcal{O}}(0)-\bar{\mathcal{O}}(t)\,,
\end{equation}
which is true for all times $t>0$, and which constitutes our first main result. 

Equation~\eqref{eq:main1} is a rigorous relationship between the mutual information and the change of the OTOC. In scrambling dynamics, $\bar{\mathcal{O}}(0)-\bar{\mathcal{O}}(t)$ is a monotonically growing function, and Eq.~\eqref{eq:main1} asserts that also $\mathcal{I}(t) $ has to be growing. This is consistent with intuitive understanding of ``scrambling'', which should be equivalent to the growth of entanglement between $A$ and $B$. We will now continue the analysis by illustrating Eq.~\eqref{eq:main1} with two important examples, before we discuss the thermodynamic significance of $\mathcal{I}(t)$.
	
\subsection{Information scrambling in experimentally relevant systems}

\subsubsection{Verified quantum information scrambling in ion traps}

The experimental verification of quantum information scrambling \cite{landsman2019verified} was conducted with a 7-qubit fully-connected quantum computer with a family of 3-qubit entangling unitaries $U(t)$\footnote{The exact and rather lengthy expressions for $U(t)$ can be found in the methods section of Ref.~\cite{landsman2019verified}.}. These entangling (scrambling) unitaries were  constructed from a combination of 1-qubit and 2-qubit gates. Due to the experimental specifics, the observables $O_A$ and $O_B$ had to be of special form. Therefore, Ref.~\cite{landsman2019verified} considered a modified version of the OTOC, namely
\begin{equation}
\mathcal{MO}(t)=\sum_{\phi, O_{\mathrm{p}}}\la O_{1}^{\dagger} O_{\mathrm{P}}^{\dagger}(t) O_{1} O_{\mathrm{P}}(t)\ra,
\label{4}
\end{equation}
where $O_{1} \equiv \ket{\psi}\bra{\phi}$ acts on the first qubit, $\ket{\psi}$ denotes the state of the qubit, and $\ket{\phi}$ is the teleported state (the last qubit of the experiment). Moreover, $O_{\mathrm{P}}(t)$ are Pauli matrices evolved by a scrambling unitary in the Heisenberg picture, and the average is taken over all Pauli matrices and state vectors $\ket{\phi}$. 

It is relatively easy to see that $\varDelta \mathcal{MO}$ $\simeq$ $\varDelta \mathcal{O}$ at times close to zero, since $O_{P}(t)$ has a simple form (in terms of operator complexity). However, in general we have $\varDelta \mathcal{MO}\leq\varDelta \mathcal{O}$, since the specific average taken in Ref.\cite{landsman2019verified} to compute $\mathcal{MO}$ is only an approximate 1-design, which does not capture the complex dynamics of $O_{P}(t)$ as the operator spreads to the other support (becomes non-local) with time. 

\begin{figure}
\centering
\includegraphics[width=.85\textwidth]{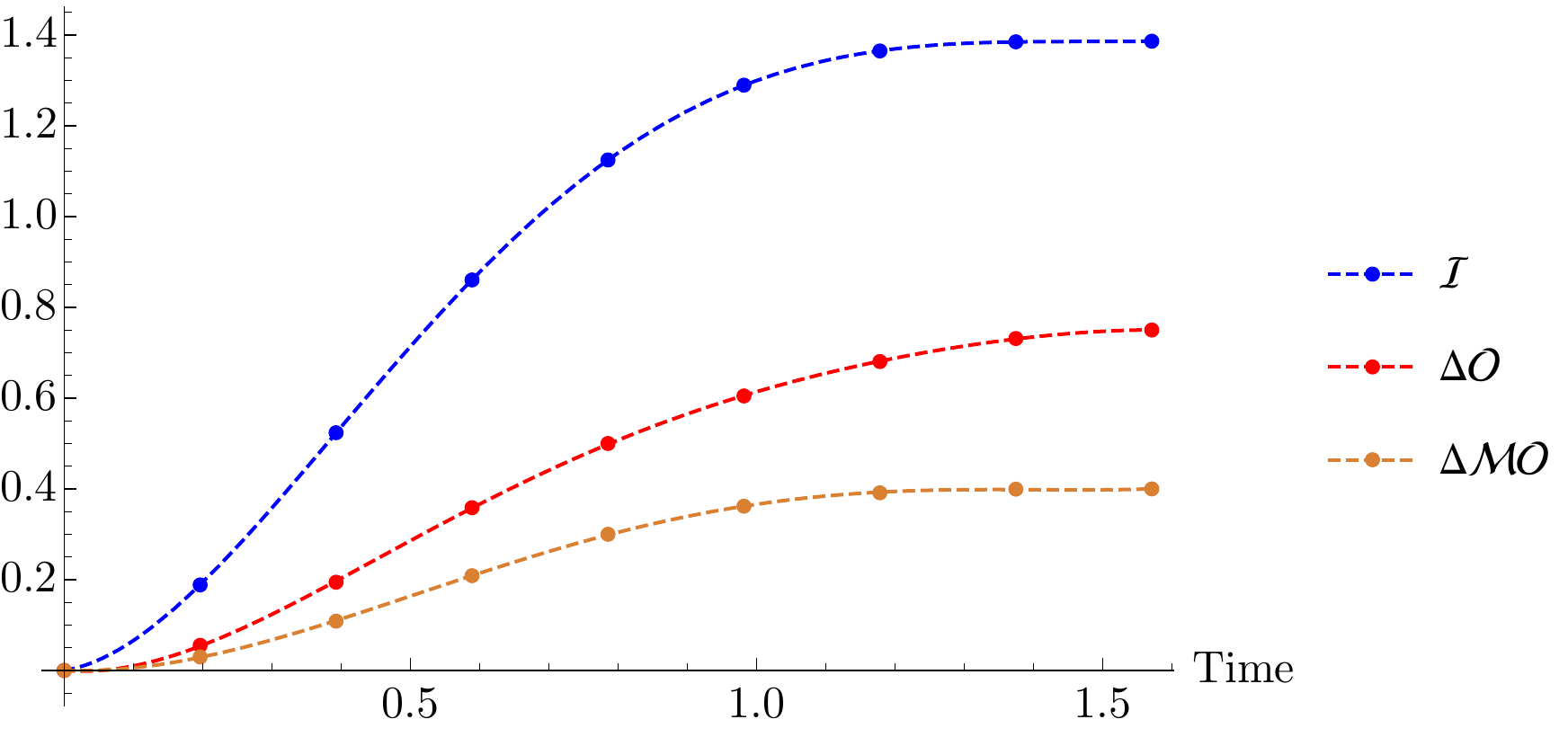}
\caption{Mutual information $\mathcal{I}$ (blue, top line), change of the OTOC $\varDelta \mathcal{O}(t)$ (red, middle line), and change of the modified OTOC $\varDelta \mathcal{MO}$ (orange, bottom line) as function of time.}
	\label{fig:Monroe}%
\end{figure}

In Fig.~\ref{fig:Monroe} we plot the mutual information \eqref{eq:mut_info}, together with $\varDelta \mathcal{O}$ \eqref{eq:main1} and $\varDelta \mathcal{MO}$ \eqref{4} for the scrambling dynamics of Ref.~\cite{landsman2019verified}.  As in Ref.~\cite{landsman2019verified}, $B=\{2,3,4,5,6,7\}$ refers to the qubits in the experiment, and $A=\{1\}$ is the first qubit. We observe that all quantities are montonically increasing functions of time $t$.

Furthermore, $\mathcal{I}(0)=0$ indicates the absence of any scrambling in the system, while $\mathcal{I}(t)= 2\ln(2)$ indicates maximal scrambling: the maximum value of the information that B can know about A (or that A can know about B) is reached. This accentuates an important advantage of $\mathcal{I}$ as a measure of scrambling over $\mathcal{MO}$. In general, $\mathcal{I}_\mrm{max}(t_*)=\mi {d_A,d_B}\ln(2)$ is the maximum value of $\mathcal{I}$ at the scrambling time, $t_*$, whereas  the maximal value of $\mathcal{MO}$ depends on the specifics of the performed experiment.

\subsubsection{Quantum chaotic dynamics -- the SYK model}

As a second example to illustrate Eq.~\eqref{eq:main1} we choose the Sachdev-Ye-Kitaev (SYK) model \cite{sachdev1993gapless,maldacena2016remarks,Rosenhaus2019}. This is an exactly solvable, chaotic many-body system consisting of $N$ interacting Majorana fermions with random interactions between $q$ of these fermions ($q$ taken as an even number). The SYK model has found important applications,  for instance, as a quantum gravity model of a $1+1$-dimensional black hole \cite{Rosenhaus2019} in the limit of large $N$.
\begin{figure}
\centering
\includegraphics[width=.85\textwidth]{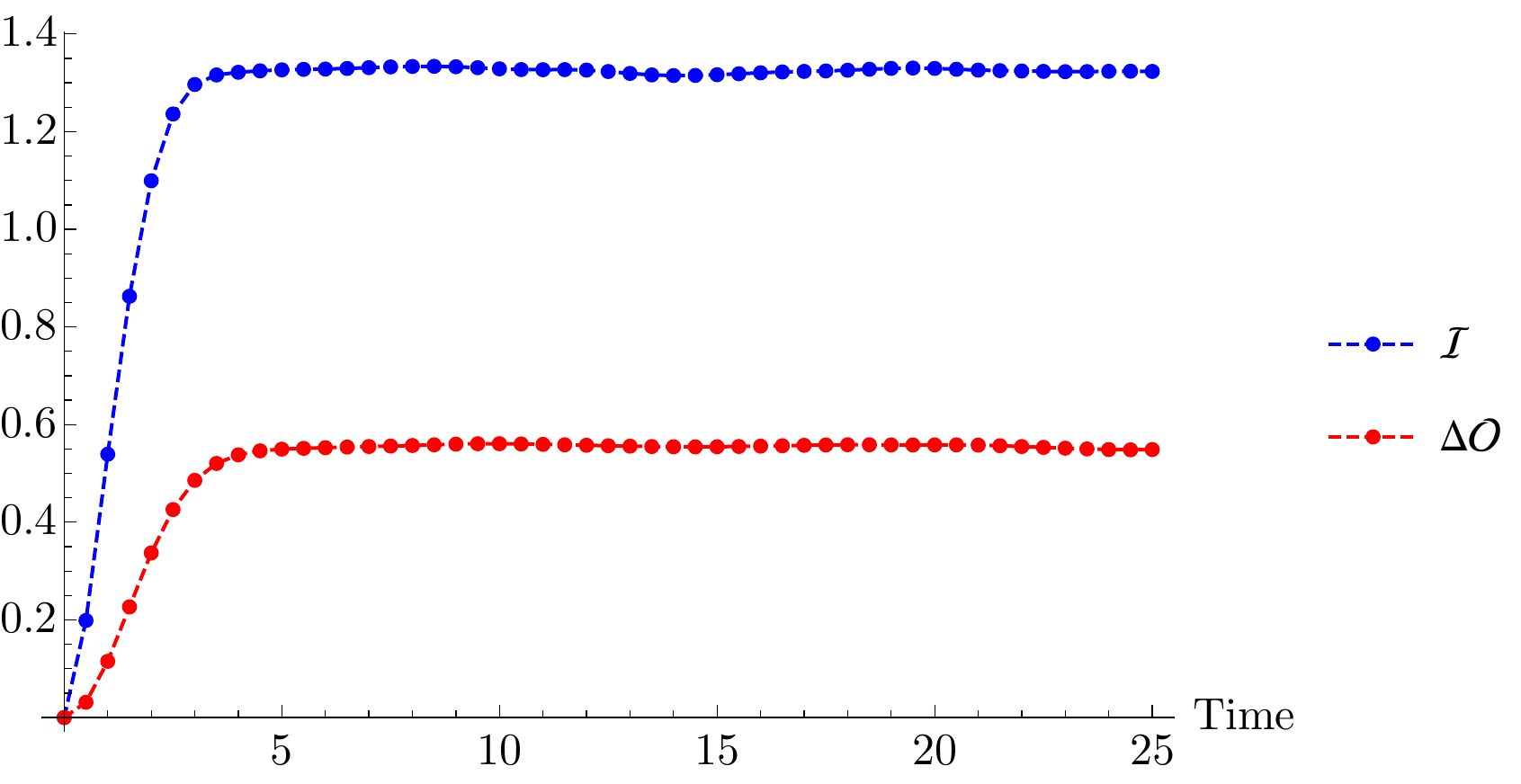}
\caption{Mutual information $\mc{I}(t)$ (blue, upper line) and $\varDelta \mathcal{O}$ (red, lower line) for the SYK model with $N=10$ Majorana fermions, $q=4$, $J^{2}$=2, and averaged over 300 realizations, where in each realization we generate a new Hamiltonian with different random interaction terms.
	\label{fig:SYK}}%
\end{figure}

The Hamiltonian can be written as
\begin{equation}
H=(i)^{\frac{q}{2}} \sum_{1 \leq i_{1}<i_{2}<\cdots<i_{q} \leq N} J_{i_{1} i_{2} \cdots i_{q}} \psi_{i_{1}} \psi_{i_{2}} \cdots \psi_{i_{q}},
\label{ham}
\end{equation}
where $J_{i_{1} i_{2} \cdots i_{q}}$ are real independent random variables with values drawn from a Gaussian distribution with mean $\left\langle J_{i_{1} \cdots i_{q}}\right\rangle=0$ and variance $\left\langle J_{i_{1} \cdots i_{q}}^{2}\right\rangle=J^{2}(q-1) !/N^{q-1}$, the parameter $J$ (in the variance) sets the scale of the Hamiltonian. Further,  $\psi_{i}$ are Majorana field operators for $i\in \{ 1,\dots, N\}$.

Remarkably, the problem can be mapped from interacting Majorana fermions to interacting qubits with random interaction terms  by using the Jordan-Wigner transformation \cite{Batista2001,cotler2019spectral}. For the exact mapping between interacting Majorana fermions and interacting spins, we follow the notation of Ref.~\cite{cotler2019spectral}, to get 
\begin{equation}
\psi_{2 j}=\frac{1}{\sqrt{2}}\left(\prod_{i=1}^{N / 2-1} \sigma_{i}^{z}\right) \sigma_{N / 2}^{y}, \quad \psi_{2 j-1}=\frac{1}{\sqrt{2}}\left(\prod_{i=1}^{N / 2-1} \sigma_{i}^{z}\right) \sigma_{N / 2}^{x},
\label{syk}
\end{equation}
such that $\forall i,j\in \{ 1,\dots, N\}$ we have $\left\{\psi_{i}, \psi_{j}\right\}=\delta_{i j}$. Notice that \myeqref{syk} demonstrates that $N$ Majorana fermions can be represented by a string of $N/2$ Pauli operators.

For the present purposes, we choose the first qubit to be subsystem $A$ and the complement of $A$  constitutes subsystem $B$. In Fig.~\ref{fig:SYK} we plot of the mutual information and together with $\varDelta \mathcal{O}$,  for a Hamiltonian with $N=10$ Majorana fermions, of which any set of $q=4$ fermions is interacting. Hence, we have a total of 210 terms in the Hamiltonian~\eqref{ham}. 

We observe that both $\varDelta \mathcal{O}$ and $\mathcal{I}$ are montonically increasing functions with time. Moreover, we see that $\varDelta \mathcal{O}$ is always upper bounded by the mutual information, and $\mathcal{I}$ reaches a maximum value slightly lower than $2\ln(2)$ given that in each realization we have small oscillations due to finite size effects. Note that for large $N$ the small oscillations (recurrences) will die out as the system becomes strongly chaotic. 

In conclusion, Eq.~\eqref{eq:main1} establishes a rigorous lower bound on the quantum mutual information and the Out-of-Time-Order Correlator. Thus, from a theoretical point of view the quantum mutual information has the same appealing properties as the OTOC in characterizing information scrambling. However, the mutual information has the added benefit that it is also a well-studied quantity in quantum thermodynamics \cite{Deffner2019book}, and it can be closely related to irreversible entropy production.

\section{Stochastic entropy production in quantum scrambling}
\label{sec3}

Quantum information scrambling is an inherently \emph{dynamical} phenomenon. Especially in chaotic quantum systems it is, thus, important to understand the rate with which information is lost to local observation \cite{Roberts2016}. Moreover, from a thermodynamic point of view, it appears appealing to relate the rate with which the quantum mutual information, $\mc{I}(t)$, grows to the \emph{local} entropy production in subsystems $A$ and $B$. Therefore, motivated by analyses of the rate of information production \cite{nikolov1994limitation,brody1995upper,Deffner2020}, we now seek to upper bound the rate of change of $\mc{I}(t)$ in terms of the thermodynamic resources (locally) consumed while scrambling information.

\subsection{Continuous quantum systems}

\begin{figure}
\centering
\includegraphics[width=0.85\textwidth]{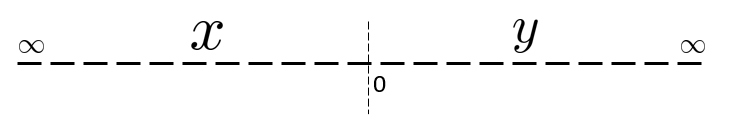}
\caption{Sketch of the two subsystems $A$ and $B$ as represented in continuous variables. \label{fig:var}}
\end{figure}

We start by considering the continuity equation for $S$ as expressed in continuous variables
\begin{equation}
\partial_{t} \rho_{S}(x,y;t)=-\boldsymbol{\nabla} \cdot \boldsymbol{j}_{S}(x,y;t)\,.
\end{equation}
Here, $\rho_{S}(x,y;t)=\bra{x y}\rho_{S}(t)\ket{x y}$ is the density function of the system evaluated in variables $x$ and $y$. Without loss of generality and for the sake of simplicity, we choose $x$ and $y$ as the coordinates in which $\rho_A$ and $\rho_B$ are diagonal, respectively, cf. Fig.~\ref{fig:var}, and $\boldsymbol{j}_{S}(x,y;t)$ denotes the probability current. Note that this choice is made purely out of mathematical convenience with the sole purpose to be able to relate the rate of change of the mutual information to stochastic entropy production: quantities that are basis independent (see below).  It has proven useful in quantum stochastic dynamics to analyze such abstract and ``non-experimental'' quantities to gain insight into the universal behavior, for instance see also Refs.~\cite{Deffner2013EPL,Micadei2020}. From an experimental point of view this choice would be highly impractical, since it requires the instantaneous diagonalization of $\rho_A$ and $\rho_B$. However, rephrasing the current treatment in a more general choice of coordinates, would require to write all expressions in terms of four variables (instead of only two) to account for the off-diagonal terms. For the present purposes we expect no additional physical insight, and such a choice would only make the mathematical expressions messier.

The corresponding, \emph{local} continuity equations are obtained by tracing out the corresponding other subsystem. In particular, we have  $\int dx\,  \rho_{S}(x,y;t)=\rho_{B}(x;t)$ and $\int dy\,  \rho_{S}(x,y;t)=\rho_{A}(y;t)$. Therefore, we can write
\begin{equation}
\partial_{t} \rho_{A}(x;t)=-\boldsymbol{\nabla}_x\cdot\boldsymbol{j}_{A}(x;t)+{j}_{S}(x,0;t).
\label{12}
\end{equation}
and
\begin{equation}
\partial_{t} \rho_{B}(y;t) =-\boldsymbol{\nabla}_y\cdot\boldsymbol{j}_{B}(y;t)-{j}_{S}(0,y;t)\,,
\label{13}
\end{equation}
where ${j}_{S}(x,0;t)$  is a boundary term that describes the influx of information from $A$ to $B$, and ${j}_{S}(0,y;t)$ is the flow from $B$ to $A$.

Now, again using that at $t=0$ subsystems $A$ and $B$ are prepared in a product state, we can write by simply taking the derivative of Eq.~\eqref{eq:mut_info}
\begin{equation}
\mathcal{\dot{I}}=-\int dx\,\left(\partial_{t} \rho_{A}\right) \ln \left(\rho_{A}\right) -\int dy\,\left(\partial_{t} \rho_{B}\right) \ln \left(\rho_{B}\right)\,.
\label{14}
\end{equation}
Employing the local continuity equations \eqref{12} and \eqref{13}, we thus have 
\begin{equation}
\begin{split}
\mathcal{\dot{I}}&=\int dx\,\left(\boldsymbol{\nabla}_x\cdot\boldsymbol{j}_{A}\right) \ln \left(\rho_{A}\right)-\int dx\,j_{S}(x,0;t)\, \ln \left(\rho_{A}\right)\\
&+\int dy\,\left(\boldsymbol{\nabla}_y\cdot\boldsymbol{j}_{B}\right) \ln \left(\rho_{B}\right) +\int dy\,j_{S}(0,y;t)\, \ln \left(\rho_{B}\right).
\end{split}
\label{15}
\end{equation}
The latter can be further simplified by partial integration, and we obtain 
\begin{equation}
\begin{split}
\mathcal{\dot{I}}=&-\int dx\,\boldsymbol{j}_{A}\cdot \boldsymbol{\nabla}_x \ln \left(\rho_{A}\right)-\int dx\,j_{S}(x,0;t)\, \ln \left(\rho_{A}\right)\\
&-\int dy\,\boldsymbol{j}_{B}\cdot \boldsymbol{\nabla}_y \ln \left(\rho_{B}\right)+\int dy\,j_{S}(0,y;t)\, \ln \left(\rho_{B}\right)\,,
\end{split}
\label{16}
\end{equation}
for which it is now easy to find upper bounds.

Using the trivial inequality, $\mathcal{\dot{I}} \leq  |\mathcal{\dot{I}}|$, and then bounding the absolute value with the Cauchy-Schwarz inequality we can write \cite{brody1995upper}
\begin{equation}
\begin{split}
\mathcal{\dot{I}} &\leq  \alpha \left(\int dx\,\frac{\boldsymbol{j}_{A}^2}{\rho_{A}}\right)^{1/2}+\gamma_{1} \left(\int dx\,\frac{\boldsymbol{j}_{S}}{\rho_{S}}(x,0;t)\right)^{1/2}\\
&+\beta \left(\int dy\,\frac{\boldsymbol{j}_{B}^2}{\rho_{B}}\right)^{1/2}+\gamma_{2}\left(\int dy\,\frac{\boldsymbol{j}_{S}}{\rho_{S}}(0,y;t)\right)^\frac{1}{2}\,,
\end{split}
\end{equation}
where we introduced the Frieden integrals \cite{nikolov1994limitation,brody1995upper},
\begin{equation}
\alpha=\int dx\,\rho_{A} \left(\boldsymbol{\nabla}_x \ln \left(\rho_{A}\right)\right)^2 \quad\text{and}\quad \beta=\int dy\,\rho_{B}\left(\boldsymbol{\nabla}_y \ln \left(\rho_{B}\right)\right)^2\,.
\end{equation}
The Frieden integral is related to the Fisher information \cite{brody1995upper}, but generally depends on the choice of variables, $x$ and $y$, and the geometry of the quantum system $S$. Similarly, we have
\begin{equation}
\gamma_{1} =\int dx\,\rho_{S}(x,0;t)\left(\ln \left(\rho_{A}\right)\right)^2 \quad\text{and}\quad \gamma_{2} =\int dy\,\rho_{S}(0,y;t)\left(\ln \left(\rho_{B}\right)\right)^2\,,
\end{equation}
which are geometric terms corresponding to the flow of information across the boundary separating $A$ and $B$. 

However, we also immediately recognize the stochastic entropy production \cite{Seifert2008,esposito2010entropy} in subsystems $A$ and $B$, which reads,
\begin{equation}
\dot{\mc{S}}_A=\int dx\,\frac{\boldsymbol{j}_{A}^2}{\rho_{A}}\quad\text{and}\quad\dot{\mc{S}}_B=\int dy\,\frac{\boldsymbol{j}_{B}^2}{\rho_{B}}\,.
\end{equation}
In conclusion, we obtain
\begin{equation}
\label{eq:bound}
\dot{\mc{I}}\leq \alpha \left(\dot{\mc{S}}_A\right)^{1/2}+\beta\left(\dot{\mc{S}}_B\right)^{1/2}+\gamma\,|\dot{\mc{S}}_E|\,,
\end{equation}
where we introduced $\gamma \,|\dot{S}_{E}|$ to denote to the \emph{exchange entropy} due to flow of information between $A$ and $B$.

Equation~\eqref{eq:bound}  provides an intuitive way to think about information scrambling, or information flow between any arbitrary partitions $A$ and $B$. The mutual information achieves a maximum if and only if the stochastic irreversible entropy productions within $A$ and $B$ as well as the entropy flow  between $A$ and $B$ vanish. So far we have only considered  scenarios where the dynamics of the quantum system is driven by information flow. A fullly thermodynamic formalism, where a system can be in contact with an information reservoir as well as the usual heat and work reservoirs \cite{deffner2013information}, will require a thoroughly developed conceptual framework which is beyond the scope of the present analysis.

\subsection{Discrete quantum systems}

The above analysis can be generalized to discrete representations of $S$. To this end, we consider the von Neumann equation describing the unitary dynamics of $\rho_{S}$
\begin{equation}
\partial_{t} \rho_{S}=-\frac{i}{\hbar}\left[H, \rho_{S}\right].
\label{18}
\end{equation}
with which the rate of change of $\mc{I}(t)$ \eqref{eq:mut_info} can be written as 
\begin{equation}
\mathcal{\dot{I}} =\frac{i}{\hbar} \left[\trace{\left[H, \rho_{S}\right] \left(\ln \left(\rho_{A}\right)\otimes I_{B}\right)}+ \trace{\left[H, \rho_{S}\right]  \left(I_{A}\otimes \ln(\rho_{B})\right)}\right]\,,
\label{19}
\end{equation}
which is mathematical a little more tedious than the continuous case. Therefore, we relegate the technical details of the derivation to the appendix.

Expressing the quantum states in Fock-Liouville space and after straightforward manipulations we again find
\begin{equation}
\mathcal{\dot{I}} \leq \mathcal{A}\, \dot{\mc{S}}_A+\mathcal{B}\, \dot{\mc{S}}_B+\mathcal{C}\,|\dot{\mc{S}}_E|\,,
\label{8}
\end{equation}
where as before $\mathcal{A}$, $\mathcal{B}$, and $\mathcal{C}$ are discrete versions of the Frieden integral \cite{nikolov1994limitation,brody1995upper}, that depend only on the geometry of the problem. Furthermore, $\dot{\mc{S}}_A$ and $\dot{\mc{S}}_B$ are the stochastic irreversible entropy production \cite{esposito2010three} in $A$ and $B$, respectively. Finally, $\dot{\mc{S}}_E$ is the entropy (or information) flow between $A$ and $B$. 

\section{Concluding remarks}

\subsection{The thermodynamic limit}

We conclude the analysis with a few remarks on thermodynamic implications. To this end, note that both quantum systems, $A$ and $B$, can be considered as open systems, for which the respective other system plays the role of an ``environment''. Imagine now that $B$ is much larger than $A$ in the sense that $B$ becomes a heat reservoir for $A$. For such a scenario and ultra-weak coupling it was shown in Ref.~\cite{esposito2010entropy} that the thermodynamic entropy, $\mc{S}_{S}(t)$, is related to the \emph{correlation entropy}, $\mc{S}_{cor}$, and we have
\begin{equation}
\mc{S}_{S}(t)=\mc{S}_{A}(t)+\mc{S}_{B}(t)+\mc{S}_{cor}(t)\,.
\label{cor}
\end{equation}
Comparing the latter with the definition of the quantum mutual information \eqref{eq:mut_info}, we immediately conclude $\mc{S}_{cor}(t)=-\mc{I}(t)$ for unitary dynamics. 

This observation indicates that the chaotic spread of quantum information is intimately related to thermalization in quantum systems -- a conclusion that can hardly be substantiated by looking only at the OTOC. Note, however, that for the above identification we had to assume that the heat reservoir $B$ is large compared to $A$, and that $B$ remains in equilibrium  at all times. More generally, one also has to account for the entropy that is produced due to the fact that the reservoir $B$ is pushed out of equilibrium through the interaction with $A$ \cite{Esposito2019,Deffner2019viewpoint}. We leave this more sophisticated analysis for a forthcoming publication \cite{Touil2020_paper2}.

\subsection{Summary}

The present analysis provides a possible starting point for the development of a comprehensive framework for the thermodynamics of information scrambling. In particular, we related the OTOC with thermodynamically relevant quantities by proving that the change of the OTOC sets a lower bound on the time-dependent quantum mutual information. This bound was demonstrated for two experimentally relevant scenarios, namely for a system of trapped ions and the SYK model. We further showed that the rate of increase of the mutual information is upper bounded by the sum of local stochastic entropy productions, and the flow of entropy between separate partitions of the quantum system.

Possible applications of our work may be sought in the study of the thermodynamics of quantum chaotic systems, and the dynamical emergence of classicality from quantum mechanics. For each of these cases, our results provide a possible route to quantify the thermodynamics resources, such as work and heat, that are consumed while information is scrambled.

\invisiblesection{Acknowledgments}
\section*{Acknowledgments}
Fruitful discussions with Nathan M. Myers and Bin Yan are gratefully acknowledged. This research was supported by grant number FQXi-RFP-1808 from the Foundational Questions Institute and Fetzer Franklin Fund, a donor advised fund of Silicon Valley Community Foundation (S.D).

\invisiblesection{Appendix}
\section{Appendix: Entropy production in Fock-Liouville space}
\renewcommand{\theequation}{B.\arabic{equation}}
\setcounter{equation}{0}  

This appendix is dedicated to the technical details that lead to Eq.~\eqref{8}. To this end, we express all quantum states in Fock-Liouville space \cite{manzano2019short}. In this formalism operators are represented as vectors and superoperators as operators, which is convenient to numerically simplify computations in Hilbert space.

We introduce the notation \cite{manzano2019short}, $\rho_{S} \equiv \left|\rho_{S}\right\rangle$ and $\left\langle \rho | \sigma\right\rangle \equiv \trace{\rho^{\dagger} \sigma}$, which defines a pre-Hilbert space and completeness is guaranteed by definition. Thus, the von Neumann equation becomes \cite{royer1991wigner},
\begin{equation}
\left|\dot{\rho}_{S}\right\rangle = W\left|\rho_{S}\right\rangle\quad\text{and}\quad W=\frac{1}{i\hbar}\,\left(H \otimes I-I \otimes H^{\boldsymbol{\top}}\right)\,.
\end{equation}
Note, that in contrast to classical stochastic dynamics that are described by rate matrices, the dynamics in Fock-Liouville space is determined by a skew Hermitian matrix, $W$.

Thus, the rate of change of $\mc{I}(t)$ \eqref{eq:mut_info} can be expressed as
\begin{equation}
\mathcal{\dot{I}} = -\sum_{m, m^{\prime}}W_{m, m^{\prime}} \rho_{m^{\prime}}\, \ln \left(\rho_{A m}^{ \prime}\right) -\sum_{m, m^{\prime}} W_{m, m^{\prime}} \rho_{m^{\prime}}\,\ln \left(\rho_{B m}^{ \prime}\right) \,,
\end{equation}
where $\rho_{A}^{ \prime}\equiv\rho_{A}\otimes I_{B}$ and $\rho_{B}^{ \prime}\equiv I_{A}\otimes \rho_{B}$. Using standard tricks from stochastic thermodynamics of adding and subtracting terms we write
\begin{equation}
\begin{split}
\mathcal{\dot{I}}& = \sum_{m, m^{\prime}} \rho_{m^{\prime}} W_{m, m^{\prime}} \ln \left(\frac{W_{m, m^{\prime}} \rho_{A m^{\prime}}^{\prime}}{W_{m^{\prime}, m} \rho_{A m}^{\prime}}\right)+\sum_{m, m^{\prime}} \rho_{m^{\prime}} W_{m, m^{\prime}} \ln \left(\frac{W_{m, m^{\prime}} \rho_{B m^{\prime}}^{\prime}}{W_{m^{\prime}, m} \rho_{B m}^{\prime}}\right)\\
&+2\sum_{m, m^{\prime}} \rho_{m^{\prime}} W_{m, m^{\prime}} \ln \left(\frac{W_{m^{\prime},m}}{W_{m,m^{\prime}} }\right)\\
& -\sum_{m^{\prime}} \ln \left(\rho_{A m^{\prime}}^{\prime}\right) \cdot\left(\sum_{m} W_{m, m^{\prime}}\right) \cdot \rho_{m^{\prime}}-\sum_{m^{\prime}} \ln \left(\rho_{B m^{\prime}}^{\prime}\right) \cdot\left(\sum_{m} W_{m, m^{\prime}}\right) \cdot \rho_{m^{\prime}}\,,
\end{split}
\label{20}
\end{equation}
which is not as involved as it looks. In particular, note that we have independent of the choice of basis, $\sum_{m,m^{\prime}} W_{m, m^{\prime}}=0$, and hence
\begin{equation}
\begin{split}
\mathcal{\dot{I}}& = \sum_{m, m^{\prime}} \rho_{m^{\prime}} W_{m, m^{\prime}} \ln \left(\frac{W_{m, m^{\prime}} \rho_{A m^{\prime}}^{\prime}}{W_{m^{\prime}, m} \rho_{A m}^{\prime}}\right)+\sum_{m, m^{\prime}} \rho_{m^{\prime}} W_{m, m^{\prime}} \ln \left(\frac{W_{m, m^{\prime}} \rho_{B m^{\prime}}^{\prime}}{W_{m^{\prime}, m} \rho_{B m}^{\prime}}\right)\\
&+2\sum_{m, m^{\prime}} \rho_{m^{\prime}} W_{m, m^{\prime}} \ln \left(\frac{W_{m^{\prime},m}}{W_{m,m^{\prime}} }\right).
\end{split}
\label{21}
\end{equation}

In complete analogy to the continuous case \eqref{16} we can now upper bound $\dot{\mc{I}}$ as
\begin{equation}
\label{eq:ent_dist}
\begin{split}
\mathcal{\dot{I}}& \leq   \sum_{m, m^{\prime}} \left|\rho_{m^{\prime}}\rho_{A m^{\prime}}^{\prime -1} \cdot \rho_{A m^{\prime}}^{\prime} W_{m, m^{\prime}} \ln \left(\frac{W_{m, m^{\prime}} \rho_{A m^{\prime}}^{\prime}}{W_{m^{\prime}, m} \rho_{A m}^{\prime}}\right)\right|\\
&+ \sum_{m, m^{\prime}} \left|\rho_{m^{\prime}}\rho_{B m^{\prime}}^{\prime -1} \cdot \rho_{B m^{\prime}}^{\prime} W_{m, m^{\prime}} \ln \left(\frac{W_{m, m^{\prime}} \rho_{B m^{\prime}}^{\prime}}{W_{m^{\prime}, m} \rho_{B m}^{\prime}}\right)\right|+\mathcal{C}\,|\dot{S}_{E}|\,,
\end{split}
\end{equation}
where we already introduced the exchange entropy production 
\begin{equation}
\mathcal{C}\,|\dot{S}_{E}|=\mathcal{A}\,\sum_{m, m^{\prime}} \left| \rho_{A m^{\prime}}^{\prime} W_{m, m^{\prime}} \ln \left(\frac{W_{m, m^{\prime}}}{W_{m^{\prime}, m}}\right)\right| +\mathcal{B}\, \sum_{m, m^{\prime}} \left| \rho_{B m^{\prime}}^{\prime} W_{m, m^{\prime}} \ln \left(\frac{W_{m, m^{\prime}}}{W_{m^{\prime}, m}}\right)\right|\,,
\end{equation}
with $\mathcal{A}=\sum_{m, m^{\prime}} \left|\rho_{m^{\prime}}\rho_{A m^{\prime}}^{\prime -1}\right|$ and $\mathcal{B}=\sum_{m, m^{\prime}} \left|\rho_{m^{\prime}}\rho_{B m^{\prime}}^{\prime -1}\right|$. Note that this is an identification only by analogy, as $W$ is not a proper rate matrix. Since $S$ is closed and evolves under unitary dynamics, the exchange entropy describes the flow of information between $A$ and $B$.

The first two terms in Eq.~\eqref{eq:ent_dist} can be further simplified to read
\begin{equation}
\mathcal{\dot{I}}\leq \mathcal{A}\,\dot{\mc{S}}_A+\mathcal{B}\,\dot{\mc{S}}_B+\mathcal{C}\,|\dot{S}_{E}|\,,
\end{equation}
where we finally introduced the \emph{local}, stochastic entropy production \cite{esposito2010three}
\begin{equation}
\dot{\mc{S}}_A= \sum_{m, m^{\prime}} \left| \rho_{A m^{\prime}}^{\prime} W_{m, m^{\prime}} \ln \left(\frac{W_{m, m^{\prime}} \rho_{A m^{\prime}}^{\prime}}{W_{m^{\prime}, m} \rho_{A m}^{\prime}}\right)\right|
\end{equation}
and
\begin{equation}
\dot{\mc{S}}_B= \sum_{m, m^{\prime}} \left| \rho_{B m^{\prime}}^{\prime} W_{m, m^{\prime}} \ln \left(\frac{W_{m, m^{\prime}} \rho_{B m^{\prime}}^{\prime}}{W_{m^{\prime}, m} \rho_{B m}^{\prime}}\right)\right|\,.
\end{equation}

\section*{References}

\bibliographystyle{iopart-num}
\bibliography{opm}

\begin{thebibliography}{10}
\bibitem [{\citenamefont {Touil}\ and\ \citenamefont
  {Deffner}()}]{Touil2020_paper2}%
  \BibitemOpen
  \bibfield  {author} {\bibinfo {author} {\bibfnamefont {A.}~\bibnamefont
  {Touil}}\ and\ \bibinfo {author} {\bibfnamefont {S.}~\bibnamefont
  {Deffner}},\ }\bibfield  {title} {\bibinfo {title} {Information scrambling
  vs. decoherence -- two competing sinks for entropy},\ }\bibinfo {note} {(to
  be published)}\BibitemShut {NoStop}%
\end{thebibliography}

\end{document}